\documentclass[11pt]{article}
\usepackage{float}
\usepackage[margin=1in]{geometry}
\usepackage{amsmath}
\usepackage{amssymb}
\usepackage{amsfonts}
\usepackage{amsthm}
\usepackage{graphicx}
\usepackage{booktabs}
\usepackage{subcaption}
\usepackage[authoryear,round]{natbib}

\usepackage{mathtools}

\newcommand{\bS}{\mathbf{S}}
\newcommand{\bX}{\mathbf{X}}
\newcommand{\bY}{\mathbf{Y}}
\newcommand{\bZ}{\mathbf{Z}}
\newcommand{\bW}{\mathbf{W}}
\newcommand{\R}{\mathbb{R}}
\newcommand{\dtime}{u}
\usepackage{xspace}
\usepackage{multirow}
\def\method{\textsc{AD-Seq-Vol}\xspace}
\def\methodft{\textsc{AD-Seq-Vol-FT}\xspace}
\def\basemethod{\textsc{AD-Seq}\xspace}
\newtheorem{remark}{Remark}
\newcommand{\colbreak}[1][]{\notag\\&#1}
\usepackage{tikz}
\usetikzlibrary{arrows.meta,fit,calc,positioning}
\usepackage{xcolor}
\definecolor{navy}{RGB}{55,73,112}

\usepackage[colorlinks,
            linkcolor=red,
            anchorcolor=blue,
            citecolor=magenta
            ]{hyperref}

\renewcommand{\colbreak}[1][]{}

\newcommand{\projroot}{../}

\IfFileExists{figures/fig_sequential_surfaces.png}{%
    \renewcommand{\projroot}{}%
    \graphicspath{{./}}%
}{%
    \graphicspath{{../}}%
}
\IfFileExists{arxiv_authors.tex}{}{}

\title{Diffusion models for dynamic volatility surface generation and data-driven hedging
\setcounter{footnote}{3}\thanks{This project is supported by a J.P. Morgan AI Faculty Research Award. The authors thank Michael Cashmore, Parisa Zehtabi, and Huining Yang for their support and helpful suggestions. We also thank Rama Cont for his valuable comments. }
}
\author{%
\setcounter{footnote}{0}%
Yinbin Han\thanks{Department of Management Science and Engineering, Stanford University. \textbf{Email:} \{yinbinha,  renyuanxu\}@stanford.edu}
\and
Jack Yuxiang Zhang\thanks{Department of Computer Science, Stanford University. \textbf{Email:} jyxzhang@stanford.edu}
\and
Manuel Torres\thanks{J.P. Morgan QTR AI Research. \textbf{Email:} \{manuel.torres, fernando.acero\}@jpmorgan.com}
\and
Fernando Acero\footnotemark[3]
\and
Renyuan Xu\textsuperscript{\ensuremath{*}}
}
\date{First version: Sep 2026; this version: Nov 2026}

\begin{document}
\maketitle
\allowdisplaybreaks

\begin{abstract}
We develop a diffusion-model framework for dynamic implied-volatility surface
generation and evaluate its economic usefulness through data-driven hedging.
The framework consists of two models. \method jointly learns the
conditional evolution of the underlying asset return and the
high-dimensional implied-volatility surface, generating adapted multi-period
scenarios by sequentially updating the realized market history.
\methodft further incorporates option-market structure through
post-training penalties for violations of static no-arbitrage conditions.
Using daily SPX option data from 2000 to 2023, we show that the proposed
models generate coherent surface trajectories while capturing both
cross-sectional and temporal dependence. \method produces fewer
and less severe static-arbitrage violations than the training data and the
GAN-based benchmark, while \methodft reduces these violations to
nearly zero. We then integrate the generated conditional scenarios into an
optimization-based hedging framework. Compared to a range of classical and data-driven benchmarks, the diffusion-based hedges maintain tracking errors near zero, substantially reduce tail risk, and exhibit particularly stable performance during the COVID-19 market disruption. These results establish Adaptive Sequential Diffusion Models as a promising class of market-consistent and economically useful financial scenario generators. Our code is available at: \url{https://github.com/yinbinhan/volatility-surface-simulation}.
\end{abstract}

\section{Introduction}

Generative models are widely used in finance to produce synthetic scenarios for
market-risk evaluation \citep{Flaig2022SGf}, tail-risk-event simulation
\citep{cont2025tail}, time-series forecasting \citep{vuletic2024fin}, and
implied-volatility surface generation \citep{vuletic2023volgan}. Most prior
applications have relied on generative adversarial networks (GANs) and
variational autoencoders (VAEs)
\citep{Wiese2020QGd,yoon2019time,ni2020conditional,vuletic2023volgan,
desai2021timevae,ning2023arbitrage}, which have known limitations: GANs
can be unstable and prone to mode collapse, whereas VAEs often
oversmooth the learned distribution. More recently,  diffusion models
\citep{ho2020denoising,song2019generative,song2020denoising,dhariwal2021diffusion}
have emerged as a powerful alternative. Their score-matching objectives avoid adversarial
optimization and have demonstrated strong performance on modeling high-dimensional, multimodal distributions.

We study whether diffusion models can capture the cross-sec\-tional and
temporal structure of financial data, focusing on implied-vol\-atil\-ity surfaces, a central input to derivative pricing, hedging and risk management \citep{gatheral2011volatility}. An implied-volatility
surface summarizes option prices across strikes and maturities 
. Each surface is a high-dimensional object
shaped by option-market dynamics and static no-arbitrage restrictions
\citep{fengler2009arbitrage,ning2023arbitrage}, and evolves jointly with
the underlying asset with substantial temporal dependence
\citep{cont2002dynamics}. A useful simulator must therefore generate
dynamically consistent trajectories conditional on realized market
history. Existing approaches primarily model the static or unconditional
distribution of individual surfaces
\citep{ackerer2020deep,ning2023arbitrage,vuletic2023volgan}. We address
this limitation using the Adaptive Sequential Diffusion Models (\basemethod) framework from
\citep{cao2026diffusion}, which learns a sequence of conditional generators
adapted to evolving market information.

Hedging provides a natural decision-based evaluation of the resulting generator, since hedge selection depends on the conditional joint evolution of the underlying asset and the volatility surface, rather than on the realism of individual surfaces in isolation. Classical model-based approaches rely on pre-specified asset-price dynamics and construct hedge positions using local sensitivities such as delta and vega \citep{Black1973Tpo,hull2022options}. Deep hedging methods instead learn hedging policies from market scenarios by minimizing a risk measure of terminal hedging error \citep{buehler2019deep,franccois2024enhancing}. The data-driven hedging framework of \cite{cont2025data} directly connects scenario generation to hedge construction: it first trains a conditional generative model on market data and then uses the generated scenarios to select hedging instruments and optimize hedge ratios.

\paragraph{Our work and main contributions.}
We propose a family of 
Adaptive Sequential Diffusion Models tailored to dynamic implied-vol\-atil\-ity surface generation. The family consists of \method, which learns the conditional evolution of the underlying asset and the implied-volatility surface, and \methodft, an arbitrage-aware variant obtained by fine-tuning the diffusion model against static no-arbitrage violations. Our main contributions are as follows.

(1). We develop \method by extending the \basemethod framework of \cite{cao2026diffusion} to high-dimensional implied-volatility surfaces. The model jointly generates the underlying asset return and the full implied-volatility surface, conditional on the recent market history. By updating this conditioning information along each generated path, it produces adapted multi-period scenarios that capture both the cross-sectional dependence across strikes and maturities and the temporal dependence of surface dynamics.

(2). We introduce \methodft, which incorporates the financial structure of option markets through arbitrage-aware fine-tuning. Specifically, we penalize violations of the calendar-spread, call-spread, and butterfly-spread constraints induced by the generated surfaces. In our empirical study of SPX options, \method already produces fewer and smaller static-arbitrage violations than the training data and the GAN-based benchmark, while \methodft reduces these violations to nearly zero.

(3). We evaluate the economic usefulness of the generated conditional scenarios through data-driven hedging. The proposed models are integrated into an optimization-based hedging procedure and compared with Black--Scholes delta hedging, delta--vega hedging, and VolGAN \citep{vuletic2023volgan}. The diffusion-based hedges keep realized tracking errors close to zero, substantially reduce tail risk relative to classical sensitivity-based hedges, and remain stable during the COVID-19 market disruption.
These results demonstrate that the surfaces generated by diffusion models are useful not only statistically, but also for downstream financial decisions.

To the best of our knowledge, this is the first work to develop a diffusion-model framework for dynamic implied-volatility surface generation and to evaluate its use in data-driven hedging.

\paragraph{Literature review and comparison.}
Our work is related to a few lines of literature.

We first review generative models for financial data, with particular attention to sequential generation and implied-volatility surfaces.

\vspace{3pt}
{\emergencystretch=1em
\noindent \underline{Generative models for financial data.}
VAE-based methods have been used for multivariate time-series and arbitrage-free option-surface simulation \citep{desai2021timevae,cai2023hybrid,liu2022time,acciaio2024time,ning2023arbitrage}. Signature-based simulators cover small-data market generation and joint spot-option simulation \citep{buehler2020generating,Buehler2020Add,Wiese2021MAS}. GAN-based methods span synthetic time series, conditional sequential generation, tail-risk simulation, trading-strategy generation, order-flow generation, and implied-volatility surfaces \citep{mirza2014conditional,esteban2017real,yoon2019time,fu2019time,Wiese2020QGd,Koshiyama2021Gan,ni2020conditional,li2020generating,vuletic2024fin,Lou2024PGg,cont2025tail,vuletic2023volgan}. Existing generators for implied-volatility surfaces rely primarily on VAE and GAN architectures \citep{ning2023arbitrage,vuletic2023volgan}; in contrast, we study diffusion models for their dynamic and history-dependent generation.\par}

\vspace{3pt}
\noindent \underline{Diffusion models for sequential data.} Score-based diffusion models have been adapted to time-series generation and forecasting \citep{rasul2021autoregressivedenoisingdiffusionmodels,lim2023regulartimeseriesgenerationusing,lim2024tsgm,yuan2024diffusionts,naiman2024utilizingimagetransformsdiffusion,yang2024survey} and to tabular and sequential financial data \citep{sattarov2023findiff,coletta2023conditional,cont2025tail}; Schr\"odinger-bridge methods provide a related stochastic-control formulation \citep{wang2021deepgenerativelearningschrodinger,Debortolietal21,hamdouche2023generativemodelingtimeseries,alouadi2025robust}. In closely related concurrent work, \citet{yang2026universal} train a single conditional diffusion model on pooled data from multiple stocks to jointly generate next-day returns and implied-volatility surface changes, demonstrating generalization to stocks unseen during training. However, the above-mentioned works do not address the adaptive generation of high-dimensional implied-volatility surface trajectories under option-market constraints. The \basemethod of \cite{cao2026diffusion} provides a general framework for conditional generation along a filtration, but its empirical scope is one-dimensional. We adapt it to the joint generation of underlying return and high-dimensional surface, introduce arbitrage-aware post-training, and evaluate the resulting scenarios through data-driven hedging.

\vspace{3pt}

\noindent\underline{Deep-learning-based implied-volatility and option simulation.} For equity-option markets, \cite{Wiese2019DHL} use GANs to generate multivariate discrete local volatilities, while \cite{Wiese2021MAS} use normalizing flows to jointly simulate spot and option markets across multiple assets. A complementary line of work uses neural-SDE approaches to model arbitrage-free joint dynamics of vanilla options \citep{Cohen2023AFN}, calibrate local stochastic volatility \citep{cuchiero2020generative}, and simulate implied-volatility surfaces across assets \citep{choudhary2024funvol}. In addition, \cite{ning2023arbitrage} combine VAEs with SDE-driven models for arbitrage-free surface generation. Furthermore, \cite{vuletic2023volgan} propose VolGAN for arbitrage-aware conditional surface generation. More recently, \cite{buehler2026sanos} construct smooth no-arbitrage option price surfaces with a non-parametric model, and \cite{buehler2026dysanos} extend the construction to a generative model. These approaches generate individual surfaces or one-step scenarios. Our models generate adapted multi-period trajectories of the underlying return and the surface by conditioning on realized history along the simulated path.

\vspace{5pt}

\noindent\underline{Data-driven hedging.} We then review learning-based approaches to hedging, distinguishing their sources of training data and their local (one-step ahead) or global (full-horizon) optimization objectives.

\vspace{3pt}

 Early work of \cite{hutchinson1994nonparametric} learns pricing functions from simulated or historical data and derives hedge ratios from their sensitivities. Deep hedging \citep{buehler2019deep}, related reinforcement learning methods \citep{cao2023gamma}, and robust deep-hedging frameworks \citep{limmer2024robust,lutkebohmert2022robust} instead learn trading policies by optimizing {\it global} criteria over a specified hedging horizon. Their numerical implementations use model-generated trajectories, such as those from the Heston model, for policy learning, with historical observations also informing calibration or evaluation in some cases.

\vspace{3pt}

 With training based entirely on historical intraday prices, \cite{mikkila2023empirical} learn a hedging policy using a {\it global} objective that maximizes expected cumulative reward across trading periods. In contrast, \cite{cont2025data} learn a conditional market generator and obtain hedge positions through one-period conditional risk minimization. We adopt this {\it local} formulation, using \method and \methodft to generate joint next-day scenarios for underlying returns and implied-volatility surfaces and optimize hedge positions with a rebalancing-cost penalty. Our generators can also supply full-horizon trajectories for hedging, although the experiments here focus on local hedging.

\section{Diffusion Models for Dynamic Implied-Volatility Surface Generation} \label{sec:diffusion}

This section develops a diffusion-model framework for generating dynamic implied-volatility surfaces. We first review the structure of implied-volatility surfaces and the static no-arbitrage constraints imposed on each surface. We then extend the framework of \basemethod of \cite{cao2026diffusion} to generate time-evolving surfaces that remain adapted to the observed market information.

\subsection{Implied volatility surfaces}
Consider an asset and let $S_t$ be the price of the underlying at time $t$, $K$ the strike, $T$ the expiration date, and $\tau=T-t>0$ the time-to-maturity. For a call option with moneyness $m=K/S_t$ and market price $c_t(m,\tau)$ in the no-arbitrage range $(S_t-Ke^{-r\tau})_+ < c_t(m,\tau) < S_t$, its Black--Scholes \textit{implied volatility} is the unique $\sigma_t(m,\tau)\in(0,\infty)$ such that
\begin{align*}
    C_{\rm BS}(S_t,K,\tau,r,\sigma_t(m,\tau))
    = S_t {N}(d_1) - K e^{-r\tau} {N}(d_2)
    = c_t(m,\tau),
\end{align*}
where
\begin{align*}
    d_1
    = \frac{-\ln m+\tau\big(r+\sigma_t^2(m,\tau)/2\big)}
    {\sigma_t(m,\tau)\sqrt{\tau}},
    d_2
    = \frac{-\ln m+\tau\big(r-\sigma_t^2(m,\tau)/2\big)}
    {\sigma_t(m,\tau)\sqrt{\tau}} .
\end{align*}
Here ${N}(\cdot)$ is the standard Gaussian c.d.f. and $r$ is the risk-free rate. The map $\sigma_t:(m,\tau)\mapsto \sigma_t(m,\tau)$ is called the implied-volatility surface at time $t$. Given the underlying price and interest-rate/dividend term structures, $\sigma_t$ is an equivalent representation of the European option price surface via the Black--Scholes formula and put--call parity \citep{gatheral2011volatility}.
Thus, admissibility of an implied volatility surface can be expressed through the static no-arbitrage properties of the corresponding option price curves, which is introduced below.

\paragraph{Static arbitrage constraints.}
We fix a grid in moneyness and time-to-maturity $(\mathbf{m},\mathbf{\tau}):=(m_i,\tau_j)_{i=1,\cdots,N_m;j=1,\cdots,N_\tau}$, with $m_i<m_{i+1}$ and $\tau_j<\tau_{j+1}$ for all $i,j$. As shown in Corollaries 4.2 and 4.3 from  \cite{davis2007range},  absence of static
arbitrage among options with strikes and maturities defined by $(\mathbf{m},\mathbf{\tau})$ is equivalent to the following three conditions:
\begin{enumerate}
    \item Absence of calendar spread arbitrage:
    \begin{eqnarray} \label{eq:arbitrage1}
        \tau_j \frac{c_t(m_i,\tau_j)-c_t(m_i,\tau_{j+1})}{\tau_{j+1}-\tau_{j}} \leq 0
    \end{eqnarray}
    for $j=1,2,\cdots, N_\tau-1$ and $i=1,\cdots, N_m$.
    \item Absence of call spread arbitrage:
    \begin{eqnarray}\label{eq:arbitrage2}
     \frac{c_t(m_{i+1},\tau_j)-c_t(m_i,\tau_{j})}{m_{i+1}-m_{i}} \leq 0
    \end{eqnarray}
    for $j=1,2,\cdots, N_\tau$ and $i=1,\cdots, N_m-1$.
    \item Absence of butterfly spread arbitrage:
    \begin{eqnarray}\label{eq:arbitrage3}
    \frac{c_t(m_{i},\tau_j)-c_t(m_{i-1},\tau_{j})}{m_{i}-m_{i-1}}    -\frac{c_t(m_{i+1},\tau_j)-c_t(m_i,\tau_{j})}{m_{i+1}-m_{i}} \leq 0
    \end{eqnarray}
    for $j=1,2,\cdots, N_\tau$ and $i=2,\cdots, N_m-1$.
\end{enumerate}
Intuitively, the calendar-spread condition \eqref{eq:arbitrage1} requires call prices to be non-decreasing in maturity, the call-spread condition \eqref{eq:arbitrage2} requires them to be non-increasing in moneyness, and the butterfly-spread condition \eqref{eq:arbitrage3} requires them to be convex in moneyness. 
Note that a non-zero positive part of the left-hand side of these inequalities
indicates the presence of static arbitrage. Following \cite{cont2023simulation}, we define the arbitrage penalty for an implied volatility surface $\sigma_t(\mathbf{m},\mathbf{\tau})$ as 
\begin{eqnarray}
\label{eq:arbitrage_loss}    L(\sigma_t(\mathbf{m},\mathbf{\tau})) = \ell_1 (\sigma_t(\mathbf{m},\mathbf{\tau})) + \ell_2(\sigma_t(\mathbf{m},\mathbf{\tau})) + \ell_3(\sigma_t(\mathbf{m},\mathbf{\tau})),
\end{eqnarray}
with
\begin{align}
    \ell_1 (\sigma_t(\mathbf{m},\mathbf{\tau})) &= \sum_{i=1}^{N_m}\sum_{j=1}^{N_\tau-1}\left( \tau_j \frac{c_t(m_i,\tau_j)-c_t(m_i,\tau_{j+1})}{\tau_{j+1}-\tau_{j}}\right)_+, \nonumber\\
    \ell_2 (\sigma_t(\mathbf{m},\mathbf{\tau})) &=\sum_{i=1}^{N_m-1}\sum_{j=1}^{N_\tau}\left( \frac{c_t(m_{i+1},\tau_j)-c_t(m_i,\tau_{j})}{m_{i+1}-m_{i}} \right)_+, \nonumber\\
    \ell_3 (\sigma_t(\mathbf{m},\mathbf{\tau})) &=\sum_{i=2}^{N_m-1}\sum_{j=1}^{N_\tau}\Big( \frac{c_t(m_{i},\tau_j)-c_t(m_{i-1},\tau_{j})}{m_{i}-m_{i-1}} \colbreak[\qquad\qquad\qquad] -\frac{c_t(m_{i+1},\tau_j)-c_t(m_i,\tau_{j})}{m_{i+1}-m_{i}}\Big)_+. \label{eq:arbitrage-decomp}
\end{align}
Static arbitrage is then equivalent to $L(\sigma_t(\mathbf{m},\mathbf{\tau})) = 0$ for each $t$. In Section~\ref{sec:experiments} we show that \method capture the static-arbitrage structure of the training data and can be further fine-tuned \citep{han2025stochastic} to reduce $L(\cdot)$.

These constraints ensure the cross-sectional admissibility of each surface at a fixed date. Dynamic surface generation further requires conditional consistency over time: generated trajectories must be adapted to realized market information and preserve the conditional distribution of future surfaces given the past. These practical considerations motivate \method introduced next.

\subsection{Adaptive sequential diffusion models for implied volatility surface generation}
We follow the Adaptive Sequential Diffusion Model (\basemethod) of \cite{cao2026diffusion}, whose original implementation is one-dimensional, and adapt it to dynamic implied-volatility surface generation. This yields two variants: \method, pretrained by denoising score matching to learn the conditional dynamics of the market, and \methodft, obtained by post-training \method with penalties for static-arbitrage violations. At each date the surface is vectorized on the moneyness-maturity grid, and the attention architecture models pairwise interactions among grid entries.

We discretize $[0,T]$ as $0=t_1<t_2<\cdots<t_H=T$ to obtain the trajectory $(\sigma_{t_h})_{h= 1}^H$, with $\sigma_{t_h}\in \mathbb{R}^{N_m N_\tau}$ after vectorization. Let $(\bW_\dtime^h)_{0\leq \dtime \leq \mathcal{T}}$ and $(\bar\bW_\dtime^h)_{0\leq \dtime \leq \mathcal{T}}$ be $2H$ independent $(N_m N_\tau)$-dimensional Brownian motions, and let $p^1(0, \cdot)$ be the density of $\sigma_{t_1}$. Fix a noise schedule $g:[0,\mathcal T]\to(0,\infty)$, where cosine or exponential are common. For $h=1$, consider the time-varying Ornstein-Uhlenbeck (OU) process defined on $\R^{N_m N_\tau}$
\begin{align}
    \mathrm{d}\bZ_\dtime^1
    =-\frac{1}{2}g(\dtime)\bZ_\dtime^1\,\mathrm{d}\dtime
+\sqrt{g(\dtime)}\,\mathrm{d}\bW_\dtime^1, \quad\bZ_0^1\sim p^1(0,\cdot). \label{eq:forward-1}
\end{align}
Let $p^1(\dtime, \cdot)$ be the marginal density of $\bZ_\dtime^1$ under \eqref{eq:forward-1}. The associated time-reversed SDE follows
\begin{align}
    \mathrm{d}\bar{\bZ}_\dtime^1
    ={}&
    g(\mathcal T-\dtime)\left(
        \frac{1}{2}\bar{\bZ}_\dtime^1
        +\nabla_z\log p^1(\mathcal T-\dtime,z)\big|_{z=\bar{\bZ}_\dtime^1}
    \right)\mathrm{d}\dtime \colbreak
    +\sqrt{g(\mathcal T-\dtime)}\,\mathrm{d}\bar{\bW}_\dtime^1, \label{eq:backward-1}
\end{align}
with $\bar{\bZ}_0^1\sim p^1(\mathcal T,\cdot)$. For $h>1$, let $ p^h_{\bf c}(0, \cdot)$ be the conditional density of $ \sigma_{t_{h}} $ given $(\sigma_{t_{1}}, \dots, \sigma_{t_{h-1}}) = {\bf c}$. The forward process for step $h>1$ is an OU process initialized from the conditional law given previously generated steps:
\begin{align}
    \mathrm{d}\bZ_\dtime^h
    =-\frac{1}{2}{g(\dtime)}\bZ_\dtime^h\,\mathrm{d}\dtime
    +{\sqrt{g(\dtime)}}\,\mathrm{d}\bW_\dtime^h, \quad \bZ_0^h \sim p_{\bar{\bZ}_{\mathcal T}^{[1:h)}}^h(0,\cdot),\label{eq:forward-2}
\end{align}
with $\bar{\bZ}_{\mathcal T}^{[1:h)} = (\bar{\bZ}_{\mathcal T}^1, \dots, \bar{\bZ}_{\mathcal T}^{h-1})$ and $p^h_{\bf c}(\dtime, \cdot)$ the marginal density of $\bZ_\dtime^h$ under \eqref{eq:forward-2}. The time-reversed SDE is
\begin{align}
    \mathrm{d}\bar{\bZ}_\dtime^h
    ={}&
    {g(\mathcal T-\dtime)}\left(
        \frac{1}{2}\bar{\bZ}_\dtime^h
        +\nabla_z\log p_{\bar{\bZ}_{\mathcal T}^{[1:h)}}^h(\mathcal T-\dtime,z)
        \big|_{z=\bar{\bZ}_\dtime^h}
    \right)\mathrm{d}\dtime \colbreak
    +{\sqrt{g(\mathcal T-\dtime)}}\,\mathrm{d}\bar{\bW}_\dtime^h, \label{eq:backward-2}
\end{align}
with $\bar{\bZ}_0^h \sim p_{\bar{\bZ}_{\mathcal T}^{[1:h)}}^h(\mathcal T,\cdot)$.

In practice, the scores in \eqref{eq:backward-1} and \eqref{eq:backward-2} are unknown and are estimated by denoising score matching \citep{vincent2011connection}. Following \cite{cao2026diffusion}, for each $h$ we train a score network $s^h$ over a class $\mathcal S^h$ by minimizing
\begin{align}
    \min_{s^h\in\mathcal S^h}
   \int_{\dtime_0}^{\mathcal T}
    \mathbb E\left[
        \left\|
        s^h(\dtime,\bZ_\dtime^h,\bZ_0^{[1:h)})
        -\frac{\alpha_\dtime\bZ_0^h-\bZ_\dtime^h}{\beta_\dtime^2}
        \right\|^2
    \right] \,\mathrm d\dtime, \label{eq:esm}
\end{align}
where $\bZ_0^{[1:h)} :=(\bZ_0^1,\ldots,\bZ_0^{h-1})$, $\dtime_0>0$ is an early-stopping time, $\alpha_\dtime=\exp\!\big(-\tfrac{1}{2}\int_0^\dtime g(s)\,\mathrm ds\big)$, $\beta_\dtime^2=1-\exp\!\big(-\int_0^\dtime g(s)\,\mathrm ds\big)$, and $\bZ_\dtime^h\mid \bZ_0^h\sim \mathcal N(\alpha_\dtime\bZ_0^h,\beta_\dtime^2 I)$ by the OU transition. Replacing the exact score with the trained $\hat s^h \in \mathcal S^h$ and the initial distribution with a standard Gaussian yields the following sampling process:
\begin{align}
    \mathrm d\widetilde{\bZ}_\dtime^h
    ={}&
    {g(\mathcal T-\dtime)}\left(
        \frac{1}{2}\widetilde{\bZ}_\dtime^h
        +\hat s^h(\mathcal T-\dtime,\widetilde{\bZ}_\dtime^h,
        \widetilde{\bZ}_{\mathcal T}^{[1:h)})
    \right)\mathrm d\dtime \colbreak
    +{\sqrt{g(\mathcal T-\dtime)}}\,\mathrm d\bar{\bW}_\dtime^h, \label{eq:backward-3}
\end{align}
where we initialize $ \widetilde{\bZ}_0^h\sim \mathcal N(0,I_{N_m N_\tau}) $. %

\paragraph{\method and its arbitrage-aware variant \methodft.}
The sampling process \eqref{eq:backward-3} supports both one-step and block generation. In one-step generation, it produces the next surface conditional on the observed history. In block generation, it is applied recursively: each generated surface is appended to the conditioning history and used to generate the next one, producing an adapted trajectory in which each surface conditions on both observed history and previously generated surfaces. We refer to this conditional sequential generator as \method.

To incorporate static no-arbitrage structure, we construct \methodft by fine-tuning the score network of \method using the diffusion fine-tuning method of \cite{han2025stochastic}, with $-L(\cdot)$ serving as the reward. Since $L(\cdot)$ vanishes when the discrete no-arbitrage constraints hold, this objective penalizes calendar-, call-, and butterfly-spread violations in the generated surfaces.

\paragraph{Suitability for implied-volatility surface generation.}
Three features of \method make it well suited to this application. First, each surface is treated as a high-dimensional vector over the grid, allowing the model to learn the cross-sectional dependence of implied volatilities. Second, the sequential conditioning on previously generated surfaces makes the sampler adapted and non-anticipative to the underlying information flow while preserving the conditional distributions of market observations \citep{cao2026diffusion}. Third, the score-based formulation provides a flexible training and fine-tuning framework that can incorporate additional admissibility criteria, such as static no-arbitrage. In Section~\ref{sec:experiments}, we evaluate \method and its no-arbitrage fine-tuned variant \methodft on implied-volatility surface generation and data-driven hedging.

\section{Data-Driven Hedging}
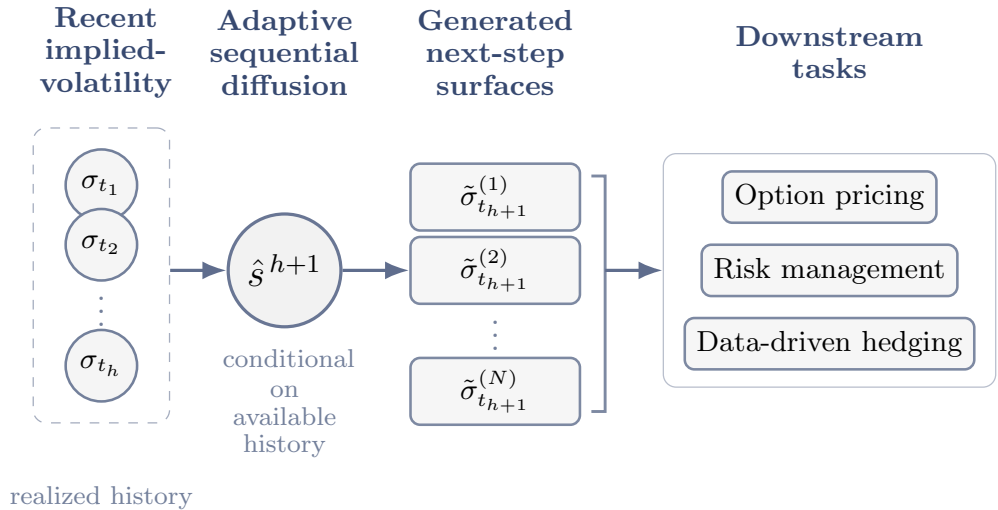
\begin{figure}[htbp]
\centering
\resizebox{0.8\linewidth}{!}{%
\begin{tikzpicture}[
    x=1cm,
    y=1cm,
    >=Latex,
    heading/.style={
        font=\footnotesize\bfseries,
        text=navy,
        align=center
    },
    history/.style={
        circle,
        draw=navy!70,
        fill=gray!8,
        line width=0.8pt,
        minimum size=0.78cm,
        inner sep=1pt,
        font=\footnotesize
    },
    scenario/.style={
        rounded corners=3pt,
        draw=navy!65,
        fill=gray!7,
        line width=0.8pt,
        minimum width=1.85cm,
        minimum height=0.50cm,
        align=center,
        font=\footnotesize
    },
    task/.style={
        rounded corners=3pt,
        draw=navy!65,
        fill=gray!7,
        line width=0.8pt,
        minimum width=1.90cm,
        minimum height=0.56cm,
        align=center,
        font=\footnotesize
    },
    mainarrow/.style={
        ->,
        line width=1.0pt,
        draw=navy!80
    },
    thin/.style={
        line width=0.9pt,
        draw=navy!65
    }
]

    \node[heading] at (0.40,2.80)
        {Recent\\[-1pt]implied-\\[-1pt]volatility};

    \node[history] (x1) at (0.35,1.35) {$\sigma_{t_1}$};
    \node[history] (x2) at (0.35,0.70) {$\sigma_{t_2}$};
    \node[font=\small, text=navy!70] (xdots) at (0.35,0.08) {$\vdots$};
    \node[history] (xh) at (0.35,-0.62) {$\sigma_{t_h}$};

    \node[
        draw=navy!45,
        dashed,
        rounded corners=4pt,
        inner xsep=0.34cm,
        inner ysep=0.22cm,
        fit=(x1)(x2)(xdots)(xh)
    ] (histbox) {};

    \node[
        font=\scriptsize,
        text=navy!70,
        anchor=north,
        align=center
    ] at ($(histbox.south)+(0,-0.55)$)
        { realized history};

    \node[heading] at (2.35,2.80)
        {Adaptive\\[-1pt]sequential\\[-1pt]diffusion};

    \node[
        circle,
        draw=navy!75,
        fill=gray!10,
        line width=1pt,
        minimum size=1.20cm,
        align=center,
        font=\large
    ] (score) at (2.35,0.42) {$\hat s^{\,h+1}$};

    \node[
        font=\scriptsize,
        text=navy!70,
        align=center,
        text width=1.9cm,
        anchor=north
    ] at (2.35,-0.30)
        {conditional on\\[-1pt]available\\[-1pt]history};

    \node[heading] at (4.65,2.80)
        {Generated\\[-1pt]next-step\\[-1pt]surfaces};

    \node[scenario] (z1) at (4.65,1.22) {$\tilde{\sigma}_{t_{h+1}}^{(1)}$};
    \node[scenario] (z2) at (4.65,0.42) {$\tilde{\sigma}_{t_{h+1}}^{(2)}$};
    \node[font=\small, text=navy!70] (zdots) at (4.65,-0.18) {$\vdots$};
    \node[scenario] (zN) at (4.65,-0.90) {$\tilde{\sigma}_{t_{h+1}}^{(N)}$};

    \coordinate (brtop) at (5.70,1.45);
    \coordinate (brmid) at (5.85,0.42);
    \coordinate (brbot) at (5.70,-1.12);

    \draw[thin] (brtop) -- (brmid |- brtop) -- (brmid |- brbot) -- (brbot);

    \node[heading] at (8.30,2.80)
        {Downstream\\[-1pt]tasks};

    \node[task] (pricing) at (8.30,1.22) {Option pricing};
    \node[task] (risk)    at (8.30,0.42) {Risk management};
    \node[task] (hedge)   at (8.30,-0.38) {Data-driven hedging};

    \node[
        draw=navy!35,
        rounded corners=4pt,
        inner xsep=0.22cm,
        inner ysep=0.18cm,
        fit=(pricing)(risk)(hedge)
    ] (taskbox) {};

    \coordinate (histout) at ($(histbox.east |- score.west)$);
    \draw[mainarrow] (histout) -- (score.west);
    \draw[mainarrow] (score.east) -- (z2.west);
    \draw[mainarrow] (brmid) -- (taskbox.west);

\end{tikzpicture}%
}
\vspace{-2pt}
\caption{Adaptive sequential diffusion for dynamic implied-volatility surface generation.}
\label{fig:iv_surface_generation}
\vspace{-8pt}
\end{figure}

Dynamic implied-volatility surface generation can serve as a conditional scenario generator for a broad range of downstream financial tasks. Given the realized market history, the Adaptive Sequential Diffusion Model of Section~\ref{sec:diffusion} produces samples of future implied-volatility surfaces, which can be used for option pricing, forecasting, risk measurement, stress testing, and optimization-based decision making (see Figure~\ref{fig:iv_surface_generation}). For these applications, the goal is not only to generate visually realistic surfaces, but also to preserve the conditional distribution of future market states and the economic structure of the induced option prices. In this section, we focus on one important downstream application: data-driven hedging, which relies directly on conditional next-step market scenarios and is a natural fit for the \method framework.

We focus on the data-driven hedging formulation of \cite{cont2025data}. At each rebalancing date $t$, we observe a market state $\bY_t=(\bS_t,\bX_t)\in \R_+^{d_S}\times \R^{d_X}$, where $\bS_t$ denotes tradable underlying prices and $\bX_t$ denotes non-price risk factors; we take $\bX_t$ to be the vectorized implied-volatility surface, which gives $d_X=N_m N_\tau$. The target portfolio has value $V_t=f(t,\bY_t)$ and the candidate hedging instruments values $H_t^i=h_i(t,\bY_t)$ for $i \in \mathcal H$, with $f,h_i:[0,T]\times \R^{d_S + d_X}\to\mathbb R$ price functions. Given next-step scenarios $\bY_{t+\Delta t}^{(k)}$, $k=1,\dots,N$, generated from the market information up to $t$, define
\[
\begin{aligned}
    \Delta V_t^{(k)}
    &= f(t+\Delta t,\bY_{t+\Delta t}^{(k)})-f(t,\bY_t),\\
    \Delta H_t^{i,(k)}
    &= h_i(t+\Delta t,\bY_{t+\Delta t}^{(k)})-h_i(t,\bY_t).
\end{aligned}
\]
At each rebalancing date, we solve the following one-period conditional hedging problem with a rebalancing-cost penalty:
\begin{align}
    \min_{A_t\in\mathbb R,\,\phi_t\in\mathbb R^{|\mathcal H|}}
    &\frac{1}{N}\sum_{k=1}^N
    \left(
        \Delta V_t^{(k)}
        - A_t
        - \sum_{i\in\mathcal H}\phi_t^i\Delta H_t^{i,(k)}
    \right)^2 \colbreak
    + \alpha g_0\sum_{i\in\mathcal H}
    c_t^i|\phi_t^i-\phi_{t-\Delta t}^i|. \label{eq:lasso}
\end{align}
Here, $|\mathcal H|$ is the number of hedging instruments, $\phi_t^i$ denotes the number of units held in instrument $i$ over $[t,t+\Delta t]$, and $A_t$ denotes an intercept. The fitted coefficients $\phi_t^i$ and $A_t$ implement hedging by making the target portfolio change $\Delta V_t^{(k)}$ close to the hedge-instrument change $A_t+\sum_{i\in\mathcal H}\phi_t^i\Delta H_t^{i,(k)}$ across scenarios \citep{cont2025data}. The first term measures one-step tracking error across generated scenarios, and can be interpreted as a discrete counterpart of the local risk-minimization criterion of \cite{follmer1991hedging} and \cite{schweizer2001guided}, evaluated under the generator-implied conditional distribution rather than under a pre-specified asset-price model. The second term penalizes rebalancing costs: the quantity $|\phi_t^i-\phi_{t-\Delta t}^i|$ is the change in the position of instrument $i$ and serves as a linear proxy for the transaction cost incurred when rebalancing. Here, $c_t^i$ is the per-unit transaction cost of instrument $i$ at time $t$, $g_0$ is the initial gross position, and $\alpha\ge0$ controls the strength of regularization. After solving \eqref{eq:lasso}, we evaluate the fitted hedge on the realized next market state $\bY_{t+\Delta t}$ and report the one-step tracking error $\varepsilon_t=\Delta V_t-A_t-\sum_{i\in\mathcal H}\phi_t^i\,\Delta H_t^i$, where $\Delta V_t$ and $\Delta H_t^i$ are the realized changes of the target portfolio and the hedging instruments. In implementation, we use the same uniform time discretization as in Section~\ref{sec:diffusion}. Since \eqref{eq:lasso} requires conditional next-step scenarios adapted to the information filtration, the adaptive sequential diffusion framework of Section~\ref{sec:diffusion} is a natural choice.

\section{Experiments}\label{sec:experiments}
In this section, we first describe the data preprocessing, model specification, and training procedure of \method for implied-volatility surfaces. We then assess the generated trajectories in terms of their qualitative surface dynamics and static-arbitrage violations, including the effect of arbitrage-aware fine-tuning. Finally, we evaluate the economic usefulness of the generated conditional scenarios through the data-driven hedging problem introduced in \eqref{eq:lasso}. We compare \method with VolGAN \citep{vuletic2023volgan} as a generative benchmark and with Black--Scholes delta and delta--vega hedges as classical hedging benchmarks. \footnote{Note that we independently implemented VolGAN, and the results we obtain are
better than those reported by the authors; nevertheless, \method exhibits more
stable and statistically better performance. See more discussions on page 6.}

\paragraph{Data preprocessing.}
Following \cite{vuletic2023volgan} and \cite{cont2025data}, we use daily SPX option quotes from OptionMetrics from January 3, 2000 to February 28, 2023, with January 3, 2000--June 16, 2018 used for training and July 1, 2018--February 28, 2023 held out for evaluation. Each daily surface is fit by a vega-weighted Nadaraya--Watson estimator with a Gaussian kernel \citep{vuletic2023volgan}, with linear interpolation in moneyness and then maturity for off-grid values. The resulting smoothed mid-price surfaces are on the moneyness grid $\{0.6,0.7,0.8,0.9,0.95,1,1.05,1.1,1.2,1.3$, $1.4\}$ and maturity grid $\{1/252,1/52,2/52,1/12,1/6$, $1/4,1/2,3/4,1\}$. The market state $\bY_t=(\bS_t,\bX_t)$ comprises the SPX log return $\bS_t$ and the log-implied-volatility surface $\bX_t$. Each training sample consists of 22 consecutive daily states, of which the first 21 form the conditioning history and the last is the one-day-ahead target; samples overlap by one trading day. The daily risk-free rate is the median rate implied by put-call parity from option mid-prices.

\paragraph{Training \method and \methodft}
We train the two models introduced in Section~\ref{sec:diffusion}, \method and its arbitrage-aware variant \methodft. The score function is parameterized by a Transformer following the implementation of \cite{cao2026diffusion} and is trained with 200 noise-injection steps and a cosine schedule $g(\cdot)$ \citep{nichol2021improved}. Optimization uses AdamW for 1000 epochs, batch size 256, initial learning rate $7\times 10^{-5}$, weight decay $10^{-2}$, and a warm-up cosine learning-rate schedule. For sampling, we use the classic denoising diffusion probabilistic model (DDPM) sampler \citep{ho2020denoising} with 200 denoising steps. \methodft is obtained by post-training the score function with the fine-tuning method of \cite{han2025stochastic}: 
We apply a rank-$8$ low-rank adaptation (LoRA) update using $-L(\cdot)$ as the reward, together with a KL penalty of strength $0.1$ relative to the pretrained model. We fine tune for 200 epochs using AdamW with batch size 32 and an initial learning rate of $10^{-4}$.

\begin{figure}[H]
  \centering
  \includegraphics[width=0.7\linewidth]{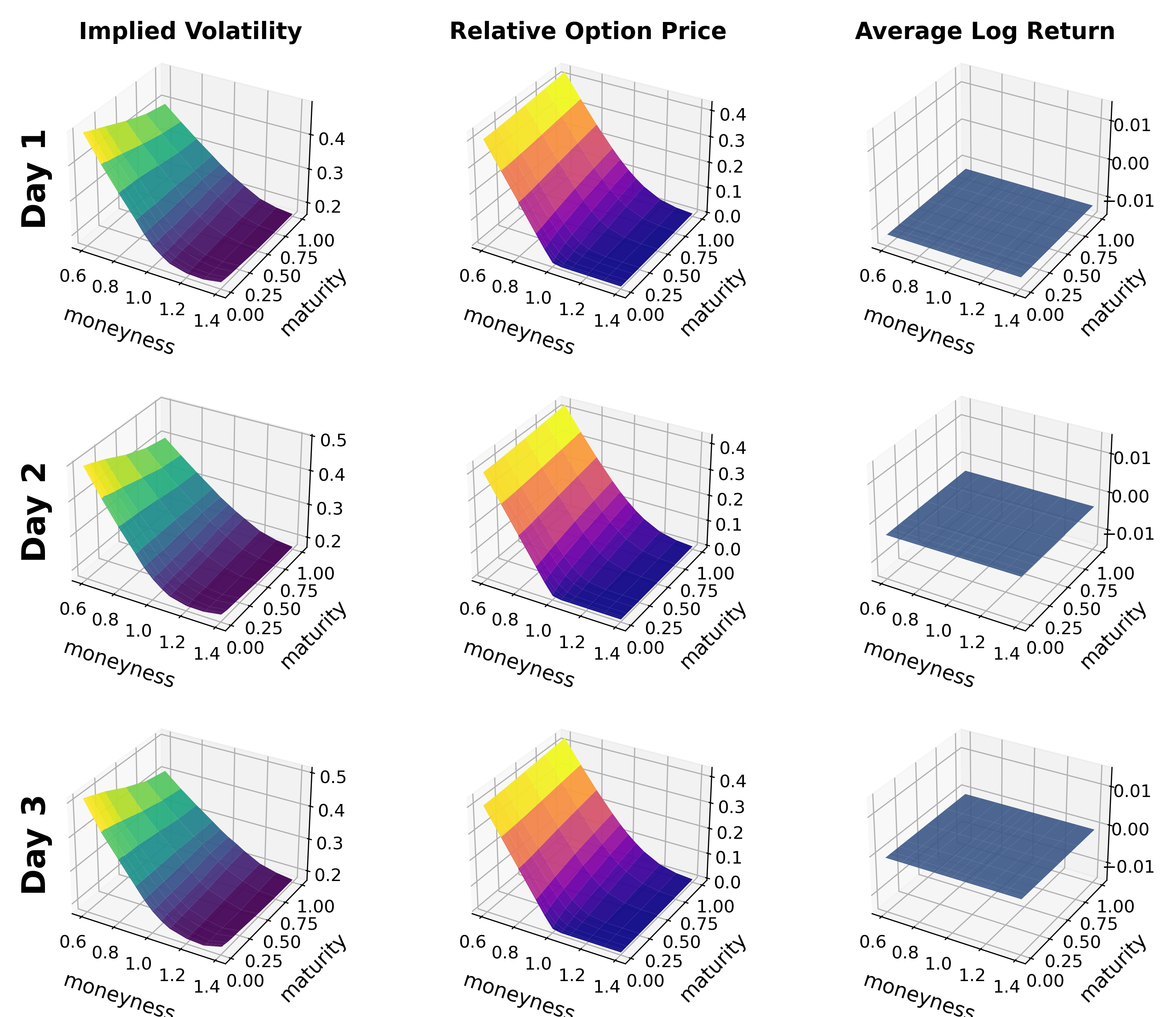}
  \caption{ A generated three-day trajectory of SPX implied-volatility surfaces (left), induced relative call-price surfaces (middle), and log-return channels (right). Each row corresponds to one day. The log-return channel at each date is a scalar daily SPX log return broadcast across the moneyness--maturity grid for visualization; its standard deviation across the grid is of order $10^{-4}$.}
  \label{fig:surfaces}
\end{figure}

\paragraph{Generated implied-volatility surfaces.}
Figure~\ref{fig:surfaces} shows a representative three-day trajectory generated by \method, with each daily surface sampled conditional on previously generated history. The trajectory exhibits the main qualitative features of SPX surfaces: monotone moneyness skew, smooth term structure across maturities, and coherent day-to-day changes in shape. The induced relative call-price surfaces are smooth and monotone in moneyness; their static-arbitrage properties are evaluated formally below. Unlike one-step generators such as VolGAN \citep{vuletic2023volgan}, \method generates multi-day trajectories recursively, using each newly generated surface to update the history that conditions the next day’s surface.

\begin{figure}[H]
  \centering
  \includegraphics[width=0.8\linewidth]{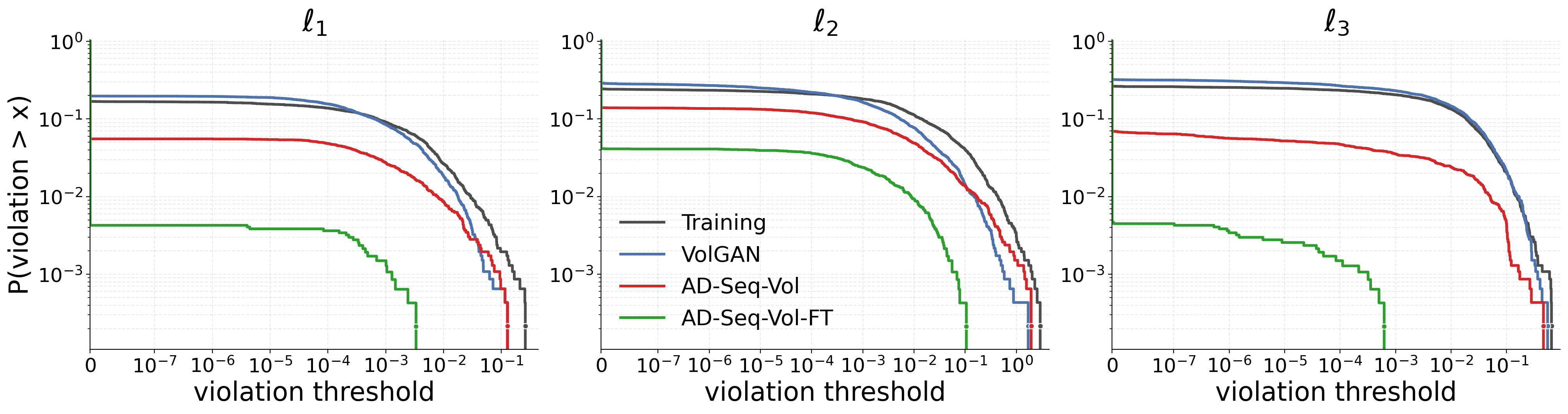}
  \caption{Fraction of surfaces with static-arbitrage violations exceeding a threshold \(x\), comparing \method, \methodft, VolGAN, and the training data. The panels show calendar-spread (\(\ell_1\), left), call-spread (\(\ell_2\), middle), and butterfly (\(\ell_3\), right) violations.}
  \label{fig:arbitrage}
\end{figure}

\paragraph{Static-arbitrage condition.}
For each surface we evaluate the arbitrage penalty $L(\cdot)$ in \eqref{eq:arbitrage_loss} and its three components $\ell_1$, $\ell_2$, $\ell_3$ from \eqref{eq:arbitrage-decomp}. Figure~\ref{fig:arbitrage} plots the fraction of surfaces whose violation exceeds a threshold $x$ for \method, \methodft, VolGAN, and the training data. On all three terms, \method has fewer violations and smaller magnitudes than the training data, and thus it does not amplify but attenuates the training-data violations. It also yields lower overall violation frequencies than VolGAN, most clearly on the calendar and butterfly constraints. Fine-tuning with \methodft  further reduces violations to near zero across all three terms, giving the lowest frequencies in Figure~\ref{fig:arbitrage}.

\paragraph{Data-driven hedging.}

We follow the experiment set-up in \cite{cont2025data} for the data-driven hedging application. For each moneyness $m_0\in\{0.75,0.8,0.9,1.1,1.2,1.25\}$ and initial date $t_1$, we construct a hedging episode in which the target position is a long straddle with strike $K=m_0S_{t_1}$. The hedge is rebalanced at each subsequent trading date until the straddle expires. The candidate hedging instruments are the SPX underlying and options on SPX with the same expiration date as the target straddle, with all strikes fixed at $t_1$: the candidate puts have strikes $0.9S_{t_1}$, $0.95S_{t_1}$, and $0.975S_{t_1}$, and the candidate calls have strikes $S_{t_1}$, $1.025S_{t_1}$, $1.05S_{t_1}$, and $1.1S_{t_1}$. The options comprising the target straddle are excluded from the candidate set. Over each rebalancing interval, the target and candidate instruments are valued at both endpoints, and their realized price changes are used to evaluate the hedging error. The transaction-cost coefficient $c_t^i$ is set to one half of the bid--ask spread of instrument $i$ at time $t$.

We compare \method with three benchmarks: Black--Scholes delta hedging, Black--Scholes delta--vega hedging, and VolGAN \citep{vuletic2023volgan}. Delta hedging uses only the SPX underlying, whereas delta--vega hedging uses the SPX underlying together with the at-the-money candidate call (strike $S_{t_1}$, sharing the target straddle's expiry); the Greeks are computed using the same Black--Scholes inputs employed for option valuation. VolGAN is implemented following \cite{vuletic2023volgan} and evaluated over the same test period as \method. For reference, Table~\ref{tab:te} also reports the published results from \cite{cont2025data}.

\begin{table}[htbp]
  \centering
  \caption{Tracking-error statistics for \(\varepsilon_t\) (USD), pooled across all \(m_0\), with and without the COVID-19 window.}
  \label{tab:te}
  \small
  \setlength{\tabcolsep}{5pt}
  \resizebox{\columnwidth}{!}{%
  \begin{tabular}{llrrrrrr}
    \toprule
    & & \multicolumn{6}{c}{Statistics} \\
    \cmidrule(lr){3-8}
    COVID-19  & Method & Mean & Median & Std & $5\%$ VaR & $2.5\%$ VaR & $1\%$ VaR \\
    \midrule
    \multirow{6}{*}{Included}
      & Unhedged & $5.15$ & $-7.26$ & $118.95$ & $170.38$ & $230.04$ & $276.87$ \\
      & Delta & $1.40$ & $-1.78$ & $41.09$ & $30.53$ & $40.63$ & $51.46$ \\
      & Delta-vega & $0.15$ & $-0.04$ & $15.53$ & $15.37$ & $21.77$ & $29.66$ \\
      & VolGAN \citep{cont2025data} & $0.55$ & $-0.16$ & $32.98$ & $12.79$ & $23.42$ & $50.79$ \\
      & \method (Ours) & $1.04$ & $-0.03$ & $\mathbf{10.55}$ & $\mathbf{11.16}$ & $\mathbf{16.48}$ & $21.33$ \\
      & \methodft (Ours) & $1.34$ & $-0.01$ & $12.01$ & $11.74$ & $16.93$ & $\mathbf{21.11}$ \\
    \midrule
    \multirow{6}{*}{Excluded}
      & Unhedged & $2.53$ & $-7.24$ & $107.72$ & $167.78$ & $222.38$ & $264.62$ \\
      & Delta & $-2.62$ & $-1.78$ & $17.41$ & $27.39$ & $35.75$ & $46.50$ \\
      & Delta-vega & $-1.18$ & $-0.05$ & $8.20$ & $14.81$ & $21.53$ & $32.72$ \\
      & VolGAN \citep{cont2025data} & $-1.05$ & $-0.18$ & $\mathbf{8.15}$ & $\mathbf{10.55}$ & $17.32$ & $33.85$ \\
      & \method (Ours) & $0.78$ & $0.02$ & $8.33$ & $10.72$ & $16.97$ & $21.54$ \\
      & \methodft (Ours) & $1.00$ & $0.08$ & $8.55$ & $11.11$ & $\mathbf{16.41}$ & $\mathbf{20.35}$ \\
    \bottomrule
  \end{tabular}%
  }
\end{table}

At each test date, we draw $N$ conditional next-step scenarios
$\{\bY_{t+\Delta t}^{(k)}\}_{k=1}^{N}$ from the trained generator. Under
each scenario, we value the target straddle and the candidate hedging
instruments using the Black--Scholes formula. We set $N=100$ for
\method and $N=1000$ for VolGAN.\footnote{Empirically,
the hedge estimates obtained from \method stabilize with $100$ generated
scenarios, whereas those obtained from VolGAN require $1000$ scenarios to
achieve comparable stability.}
We then solve the LASSO problem
in \eqref{eq:lasso} by coordinate descent, with the regularization
parameter $\alpha$ selected according to the Akaike information
criterion, also following \cite{cont2025data}.

\begin{figure}[H]
  \centering
  \includegraphics[width=0.8\linewidth]{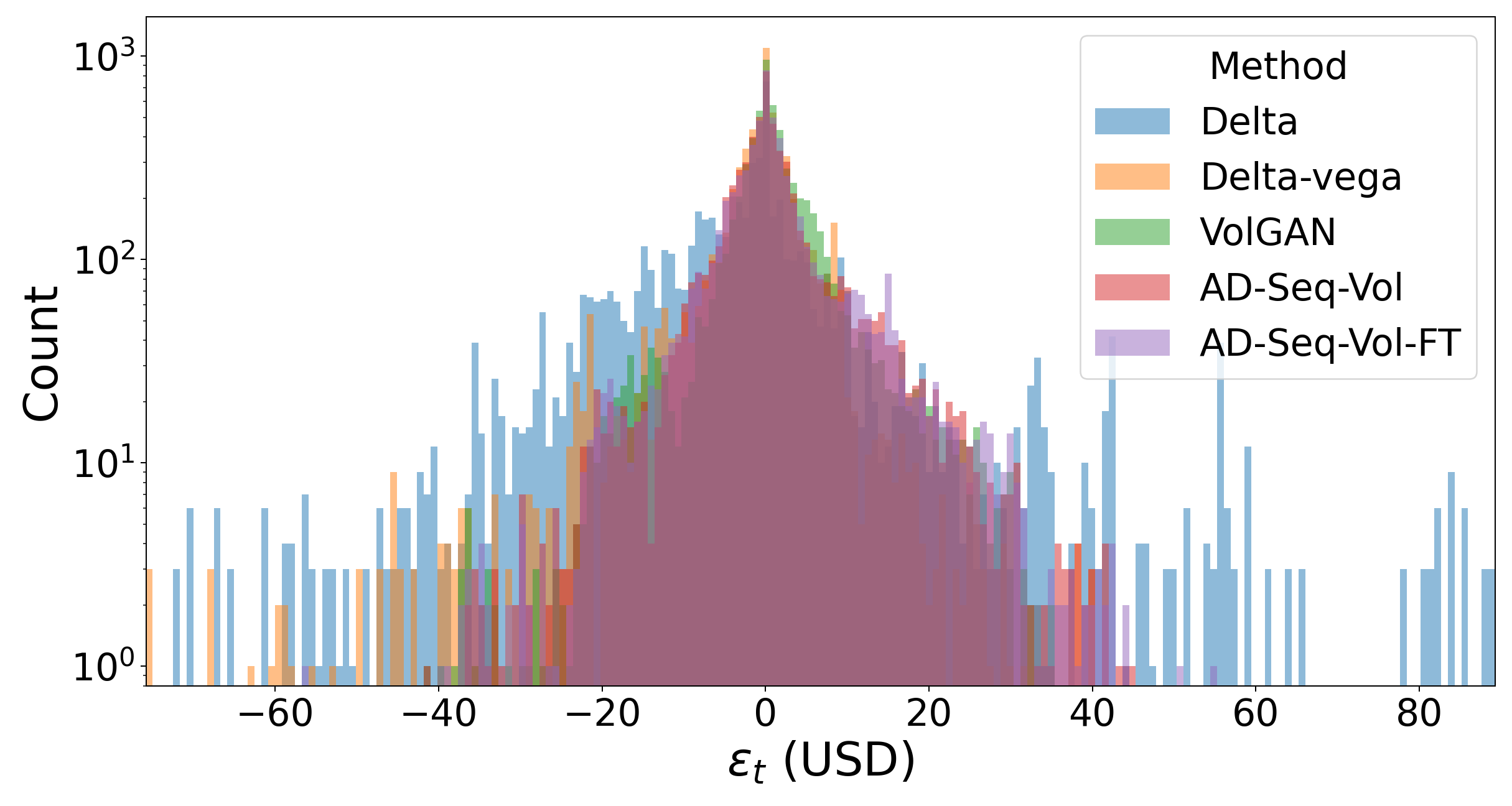}
  \caption{Tracking-error $\varepsilon_t$ distribution pooled over all $m_0$, for delta, delta--vega, VolGAN, \method, and \methodft hedging, over the full range including outliers.}
  \label{fig:te-hist}
\end{figure}

Table~\ref{tab:te} reports the realized tracking errors $\varepsilon_t$ of all hedging approaches. \method delivers the strongest and most consistent performance. Its tracking errors stay centered near zero, with medians of $-0.03$ and $0.02$ in the COVID-included and COVID-excluded windows, where the latter removes 13 February--21 July 2020. In the COVID-included window, \method attains the lowest standard deviation and the lowest $5\%$ and $2.5\%$ VaRs, with a standard deviation of $10.55$ against $32.98$ for VolGAN and $41.09$ for delta hedging. Its fine-tuned variant \methodft attains the lowest $1\%$ VaR in both windows, at $21.11$ and $20.35$, and the lowest $2.5\%$ VaR when COVID is excluded ($16.41$). This shows that the tail advantage of the diffusion-based hedges does not come from the COVID period alone. Consistent with these results, Figure~\ref{fig:te-hist} shows that the tracking errors of \method are tightly concentrated around zero, whereas the classical hedges exhibit substantially wider tails and more extreme outliers.

\begin{figure}[H]
  \centering
  \includegraphics[width=0.8\linewidth]{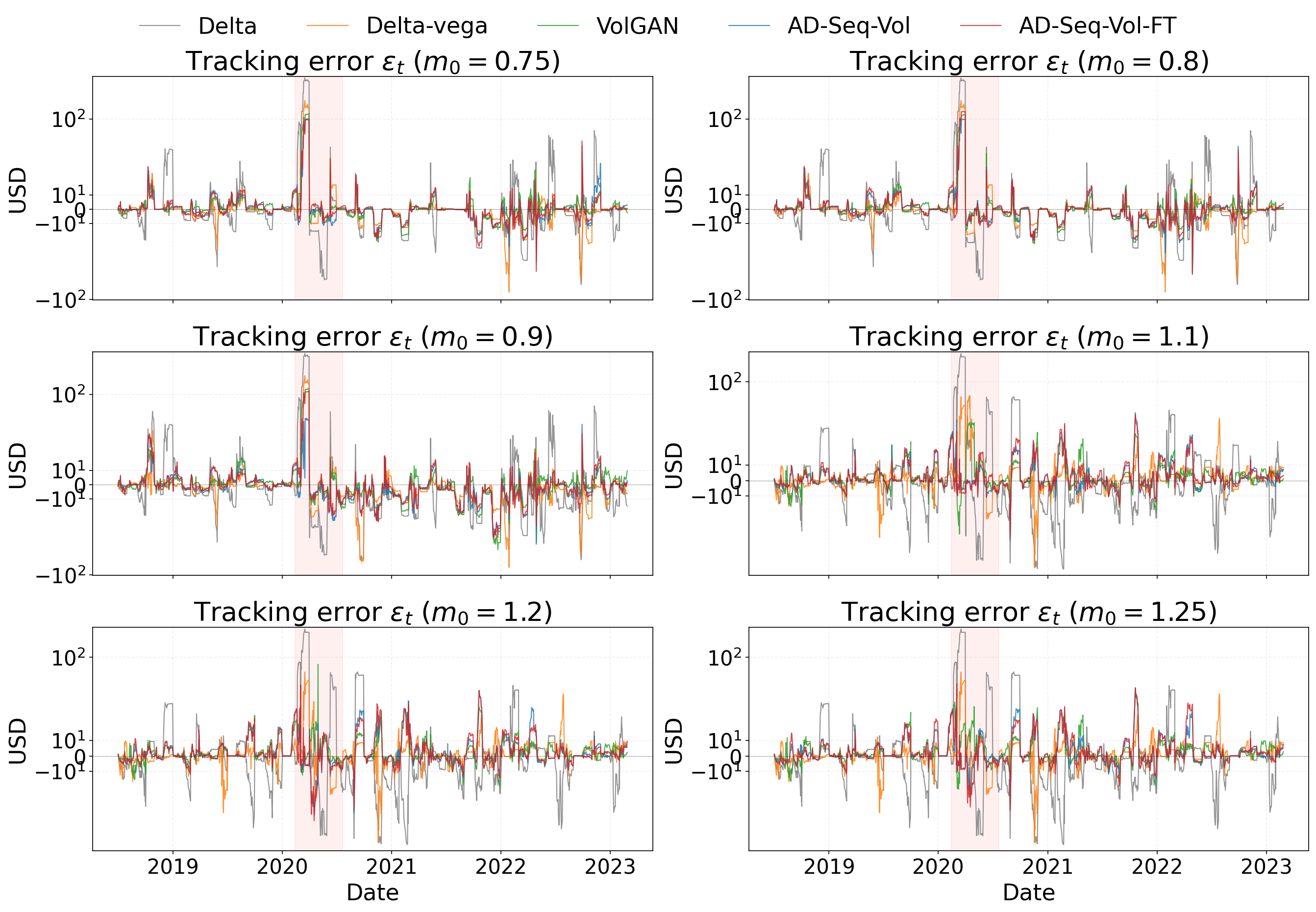}
  \vspace{-4pt}
  \caption{Tracking error \(\varepsilon_t\) over time for delta, delta--vega, VolGAN, \method, and \methodft hedging at each moneyness level \(m_0\). The shaded region marks the COVID-19 window from 13 February to 21 July 2020.}
  \label{fig:te-ts}
  \vspace{-2pt}
\end{figure}

Figure~\ref{fig:te-ts} plots \(\varepsilon_t\) at each rebalancing date for each moneyness level \(m_0\), with the COVID-19 period shaded in red. Across all panels, \method is the most stable. Tracking errors under delta and delta--vega hedging exhibit sharp spikes and prolonged dislocations during the COVID-19 shock. By contrast, those under \method and \methodft closely track one another and show only a short-lived excursion before reverting to zero. Both diffusion-based models are at least as stable as VolGAN and substantially less volatile than the classical hedging strategies.

\begin{figure}[H]
  \centering
  \includegraphics[width=0.8\linewidth]{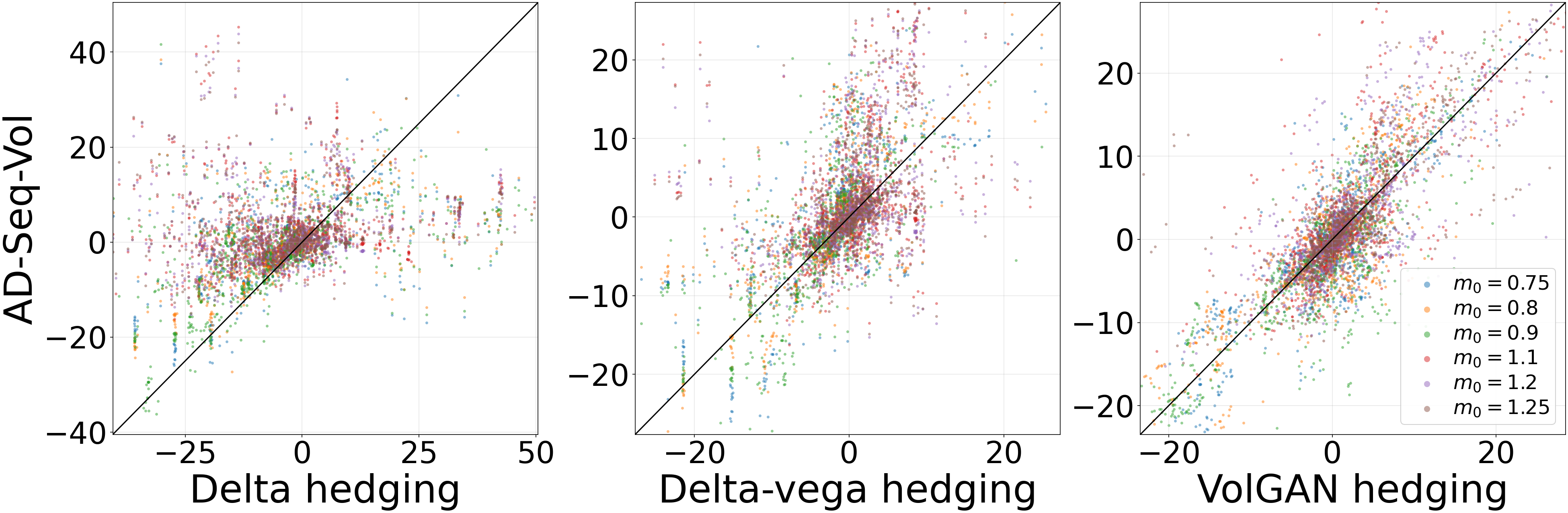}
  \vspace{-4pt}
  \caption{Tracking error of \method against delta hedging (left), delta--vega hedging (middle), and VolGAN hedging (right), colored by $m_0$. Solid line denotes $x=y$.}
  \label{fig:scatter}
  \vspace{-2pt}
\end{figure}

Figure~\ref{fig:scatter} compares \method with each benchmark over all episodes. Against delta hedging (left), the scatter is asymmetric around the diagonal: when both methods incur negative tracking errors, the magnitudes are larger under delta. Against delta--vega (middle), the observations are distributed more symmetrically around the diagonal, which shows \method performs comparably to the parametric Greek-based hedge. Against VolGAN (right), the tracking errors are concentrated around the diagonal with correlation $0.83$, indicating that the two data-driven methods agree on the bulk of hedge selection; the advantage of \method in the Std and tail-VaR statistics of Tables~\ref{tab:te} arises from a few time periods in which VolGAN produces large tracking-error excursions that \method avoids.

\vspace{10pt}
We provide a few final remarks before conclusion.
\begin{remark}{Diffusion model vs GAN-based method.} \rm
    Compared with VolGAN, \method keeps tracking errors closer to zero across all moneyness levels and substantially reduces tail risk, including during the COVID-19 market disruption. The advantage over VolGAN is most pronounced for the near-money straddles $m_0\in\{0.9,1.1\}$, whose high vega makes hedging especially sensitive to movements in at-the-money implied volatility. In the COVID-included window, for $m_0=0.9$, \method reduces the tracking-error standard deviation from $68.66$ under VolGAN to $10.49$ and the $1\%$ VaR from $80.78$ to $29.36$. For $m_0=1.1$, it reduces the $1\%$ VaR from $47.58$ to $10.26$. Our diffusion-model-based models also retain lower tail risk when the COVID-19 window is excluded.

These findings point to a broader potential advantage of diffusion models for financial applications. Existing implied-volatility generators largely model static or unconditional distributions \citep{ackerer2020deep,ning2023arbitrage}, while VolGAN generates the next-day surface from the current market state. By contrast, \method generates multi-day trajectories sequentially, conditioning each step on the evolving history. More generally, diffusion models combine a non-adversarial score-matching objective with flexible history-conditioned generation, which may offer greater stability for high-dimensional financial scenarios, particularly in stressed market regimes. The near-money results provide concrete evidence that this stability can translate into economically meaningful improvements in hedging.

\end{remark}

\begin{remark}{Our independent implementation of VolGAN.} \rm
Our implementation of VolGAN with data-driven hedging \citep{cont2025data} achieves better results than those reported by \cite{cont2025data}. Including the COVID-19 period, our implementation of VolGAN achieves a tracking-error standard deviation of \(12.10\), compared with \(32.98\) reported in \cite{cont2025data}, and a \(1\%\) VaR of \(20.49\), compared with \(50.79\). Even against this stronger implementation, \method performs better: its standard deviation is $10.55$ versus $12.10$, its $5\%$ VaR is $11.16$ versus $12.02$, and its median tracking error is $-0.03$ versus $0.31$. One possible explanation for the improvement over the published VolGAN is the number of hedging instruments selected. Our implementation of VolGAN uses $5.51$ of the eight candidate instruments on average, with a median of six, whereas \cite{cont2025data} report using only two or three options. Although a denser hedge can track the target more closely, it also requires more transactions. By comparison, \method uses only $4.73$ instruments on average, with a median of five. Thus, \method outperforms this stronger VolGAN benchmark while using a sparser hedge. 
\end{remark}

\begin{remark}{Effect of arbitrage-aware fine-tuning on hedging.} Ar\-bi\-trage-aware fine-tuning improves the tail behavior of the hedging strategy. In both windows of Table~\ref{tab:te}, \methodft attains a standard deviation and $5\%$ VaR comparable to \method and a smaller $1\%$ VaR. The improvement is most pronounced for the high-vega near-money straddles. At $m_0\in\{0.9,1.1\}$ (Table~\ref{tab:te-nearmoney}), \methodft attains the lowest $5\%$ and $2.5\%$ VaRs among all methods in both windows, reducing the $2.5\%$ VaR at $m_0=0.9$ from $22.79$ to $21.16$ with the COVID window included and from $25.52$ to $21.80$ with it excluded. Since \methodft also reduces the static-arbitrage violations of \method to near zero (Figure~\ref{fig:arbitrage}), enforcing no-arbitrage admissibility through fine-tuning improves the tail metrics  important for risk management without sacrificing hedging accuracy.
\end{remark}

\begin{table}[htbp]
  \centering
  \caption{Tracking-error $\varepsilon_t$ statistics (USD)
  for the near-money straddles $m_0\in\{0.9,1.1\}$, with the COVID-19 window
  included and excluded.}
  \label{tab:te-nearmoney}
  \small
  \setlength{\tabcolsep}{4pt}
  \resizebox{\columnwidth}{!}{%
  \begin{tabular}{cllrrrrrr}
    \toprule
    $m_0$ & COVID-19 & Method & Mean & Median & Std & $5\%$ VaR & $2.5\%$ VaR & $1\%$ VaR \\
    \midrule
    \multirow{8}{*}{$0.9$}
      & \multirow{4}{*}{Included}
        & Delta                    & $-0.09$ & $-5.61$ & $48.64$ & $32.90$ & $39.07$ & $49.53$ \\
      & & Delta-vega               & $-2.37$ & $-2.13$ & $20.43$ & $21.44$ & $28.24$ & $45.57$ \\
      & & VolGAN \citep{cont2025data} & $4.02$ & $-0.32$ & $68.66$ & $34.48$ & $55.92$ & $80.78$ \\
      & & \method (Ours)           & $-2.19$ & $-2.20$ & $\mathbf{10.49}$ & $19.05$ & $22.79$ & $\mathbf{29.36}$ \\
      & & \methodft (Ours)         & $-0.97$ & $-1.84$ & $15.09$ & $\mathbf{18.53}$ & $\mathbf{21.16}$ & $30.77$ \\
      \cmidrule(lr){2-9}
      & \multirow{4}{*}{Excluded}
        & Delta                    & $-4.76$ & $-4.95$ & $17.92$ & $32.06$ & $35.76$ & $48.19$ \\
      & & Delta-vega               & $-4.39$ & $-2.09$ & $10.03$ & $21.42$ & $28.87$ & $47.70$ \\
      & & VolGAN \citep{cont2025data} & $-2.87$ & $-0.39$ & $16.30$ & $32.88$ & $51.28$ & $73.20$ \\
      & & \method (Ours)           & $-2.63$ & $-1.92$ & $\mathbf{9.13}$ & $20.43$ & $25.52$ & $30.64$ \\
      & & \methodft (Ours)         & $-2.43$ & $-1.63$ & $9.46$ & $\mathbf{19.38}$ & $\mathbf{21.80}$ & $\mathbf{30.61}$ \\
    \midrule
    \multirow{8}{*}{$1.1$}
      & \multirow{4}{*}{Included}
        & Delta                    & $1.32$ & $-1.60$ & $33.24$ & $32.73$ & $42.84$ & $53.83$ \\
      & & Delta-vega               & $1.53$ & $0.78$ & $10.27$ & $10.13$ & $20.61$ & $24.27$ \\
      & & VolGAN \citep{cont2025data} & $-2.82$ & $-0.74$ & $13.79$ & $19.17$ & $31.41$ & $47.58$ \\
      & & \method (Ours)           & $2.41$ & $0.20$ & $8.45$ & $7.11$ & $8.50$ & $\mathbf{10.26}$ \\
      & & \methodft (Ours)         & $2.78$ & $0.60$ & $\mathbf{8.36}$ & $\mathbf{6.61}$ & $\mathbf{8.25}$ & $10.55$ \\
      \cmidrule(lr){2-9}
      & \multirow{4}{*}{Excluded}
        & Delta                    & $-2.13$ & $-1.68$ & $18.85$ & $28.48$ & $40.74$ & $54.14$ \\
      & & Delta-vega               & $0.59$ & $0.70$ & $\mathbf{7.21}$ & $8.26$ & $15.04$ & $23.67$ \\
      & & VolGAN \citep{cont2025data} & $-2.13$ & $-0.72$ & $7.84$ & $16.40$ & $24.08$ & $33.99$ \\
      & & \method (Ours)           & $2.95$ & $0.57$ & $8.48$ & $6.41$ & $7.95$ & $\mathbf{10.17}$ \\
      & & \methodft (Ours)         & $3.10$ & $0.89$ & $8.28$ & $\mathbf{6.24}$ & $\mathbf{7.57}$ & $10.41$ \\
    \bottomrule
  \end{tabular}%
  }
  \vspace{-8pt}
\end{table}

\section{Conclusion}
We developed an Adaptive Sequential Diffusion framework for dynamic
implied-volatility surface generation and evaluated the resulting market
scenarios through data-driven hedging. \method jointly models the
conditional evolution of the underlying asset return and the
high-dimensional volatility surface, while \methodft incorporates static
no-arbitrage structure through fine-tuning. In our empirical study of SPX
options, \method generates coherent, history-dependent surface trajectories
with fewer and smaller static-arbitrage violations than the training data and
the GAN-based benchmark, while \methodft reduces these violations to
nearly zero. When integrated into data-driven hedging, the diffusion-generated
scenarios yield tracking errors close to zero and improve tail-risk
performance, particularly during the COVID-19 market disruption. These
results demonstrate the potential of adaptive sequential diffusion models as
market-consistent and economically useful financial scenario generators.

Several directions remain interesting and open for future work. A natural extension is to move beyond SPX to a joint, multi-asset surface generator, spanning multiple indices and single-name options. Arbitrage-aware fine-tuning could be extended from static, per-surface conditions to dynamic no-arbitrage constraints enforced along the trajectory. Methodologically, distillation or fewer-step samplers could reduce the cost of sequential diffusion, broadening the framework's applicability to high-frequency settings. 
\bibliographystyle{plainnat}
\bibliography{\projroot reference}
\end{document}